\documentclass[conference]{IEEEtran}

\usepackage{graphicx}  

\usepackage{amsmath}   
\begin{document}

\title{On the Relationship between Transmission Power and Capacity of an Underwater
Acoustic Communication Channel}

\author{\authorblockN{Daniel E. Lucani}
\authorblockA{LIDS, MIT\\
Cambridge, Massachusetts, 02139\\
Email: dlucani@mit.edu}
\and
\authorblockN{Milica Stojanovic}
\authorblockA{MIT\\
Cambridge, Massachusetts, 02139\\
Email: millitsa@mit.edu}
\and
\authorblockN{Muriel M\'edard}
\authorblockA{LIDS, MIT\\
Cambridge, Massachusetts, 02139\\
Email: medard@mit.edu}}


%


\maketitle
\begin{abstract}

The underwater acoustic channel is characterized by a path loss that depends
not only on the transmission distance, but also on the signal
frequency. As a consequence, transmission bandwidth depends on the
transmission distance, a feature that distinguishes an underwater
acoustic system from a terrestrial radio system. The exact relationship between power, transmission band, distance and  capacity for the Gaussian noise scenario is a complicated one.
This work provides a closed-form approximate model for 1) power consumption, 2) band-edge frequency and 3) bandwidth as functions of distance and capacity required for a data link. This approximate model is obtained by numerical evaluation of analytical results which takes into account physical models of acoustic
propagation loss and ambient noise.
The closed-form approximations may become useful tools in the design and analysis of underwater acoustic networks.
\end{abstract}


%
\IEEEpeerreviewmaketitle

\section{Introduction}

With the advances in acoustic communication technology, the interest in study and experimental deployment of underwater networks has been growing \cite{partan06}. However, underwater acoustic channels impose many constraints that affect the design of wireless networks.	They are characterized by a path loss that depends on both the transmission distance and the signal frequency, a feature that distinguishes an underwater
acoustic system from a terrestrial radio system. Thus, not only the power consumption, but also the useful bandwidth depend on the transmission distance \cite{milica06}. 

	From an information theoretic perspective, both the distance between two nodes and the required capacity determine the power consumption for that link and the optimal transmission band. It is thus of interest to have a simple, closed-form expression that relates the transmission power to the desired capacity. This would enable an efficient design of both point to point links and underwater networks, eventually leading to a minimum cost overall network optimization. Thus, these expressions may be useful from both a theoretic and an engineering standpoint.
	
	In this paper, simple closed-form approximations for the power consumption and operating frequency band as functions of distance and capacity are presented. This approximate model stems from an information theoretic analysis that takes into account a physical model of acoustic propagation loss, and colored Gaussian ambient noise. It was shown in \cite{milica06} that the transmission power as a function of the distance could be well approximated by $P(l) = pl^{\gamma}$. A similar relationship was shown to exist for the operating bandwidth. The coefficients in this model were determined as functions of the required signal to noise ratio. 

	The present work extends this idea of modeling the power and bandwidth as functions of distance, but the problem is cast into a slightly different framework. Namely, instead of using the SNR as a constraint, i.e. a fixed design parameter, the desired link capacity is used as a figure of merit. In few words, this work proposes approximate models for the parameters as functions of the capacity. This resulting model is useful for a broad range of capacities and distances.
	
	The paper is organized as follows. In Section 2, a model of an underwater channel is outlined. In Section 3, a brief description of the numerical evaluation procedure is described. In Section 4, closed-form expressions for the parameters of interest are presented. Section 5 gives numerical results for different ranges of distance and capacity. Conclusions are summarized in the last section. 

\begin{figure}[t]
\centering						
\includegraphics[height=3in,width=3.5in]{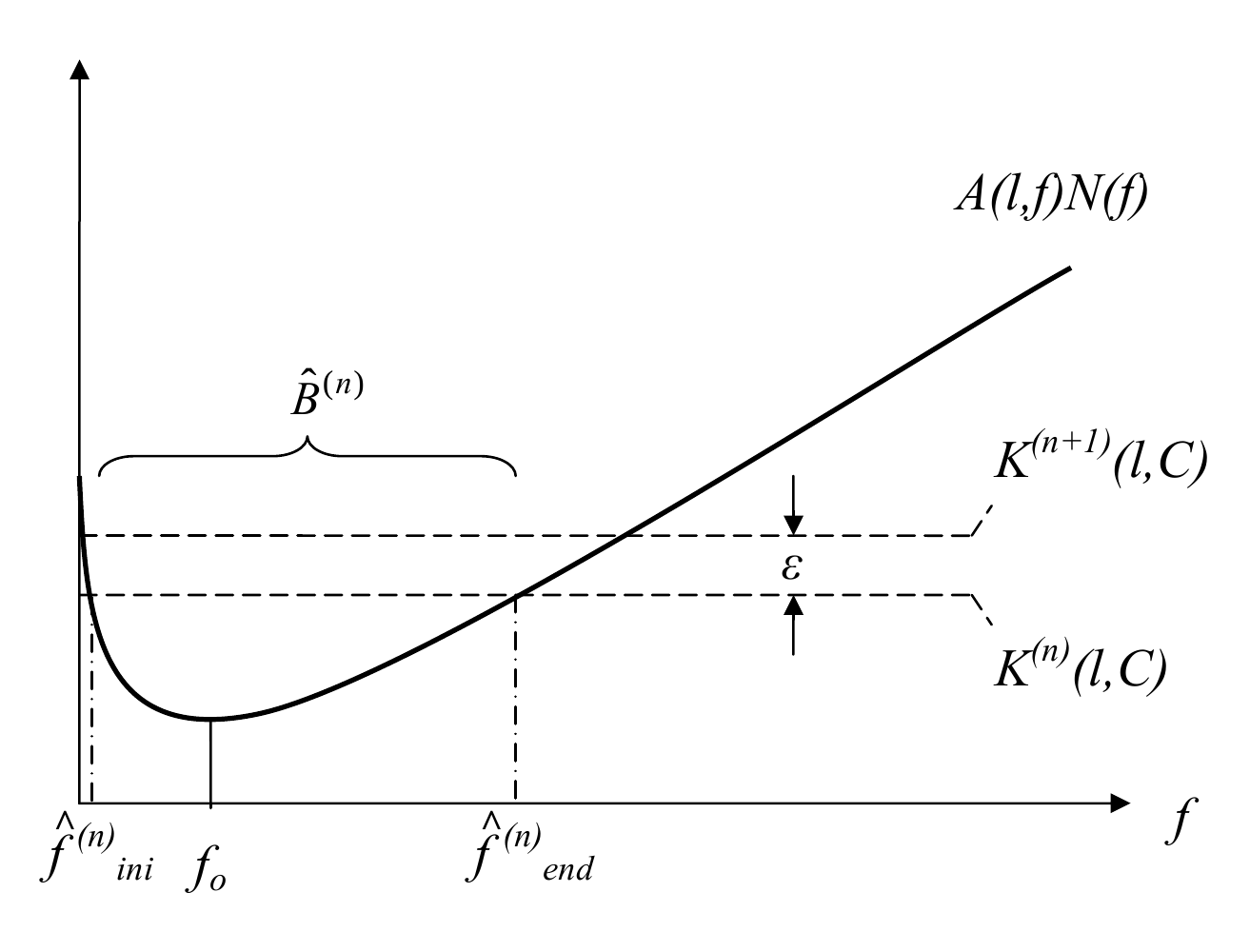}
\caption{Numerical procedure for computation of $P(l,C)$, $\hat{f}_{end}(l,C)$ and $\hat{B}(l,C)$, by incrementing $K(l,C)$ at each step by $\epsilon$ until a stopping condition is fulfilled}
\label{ANplot.tag}
\end{figure}

\begin{figure}[t]
\centering						
\includegraphics[height=3.5in,width=3.5in]{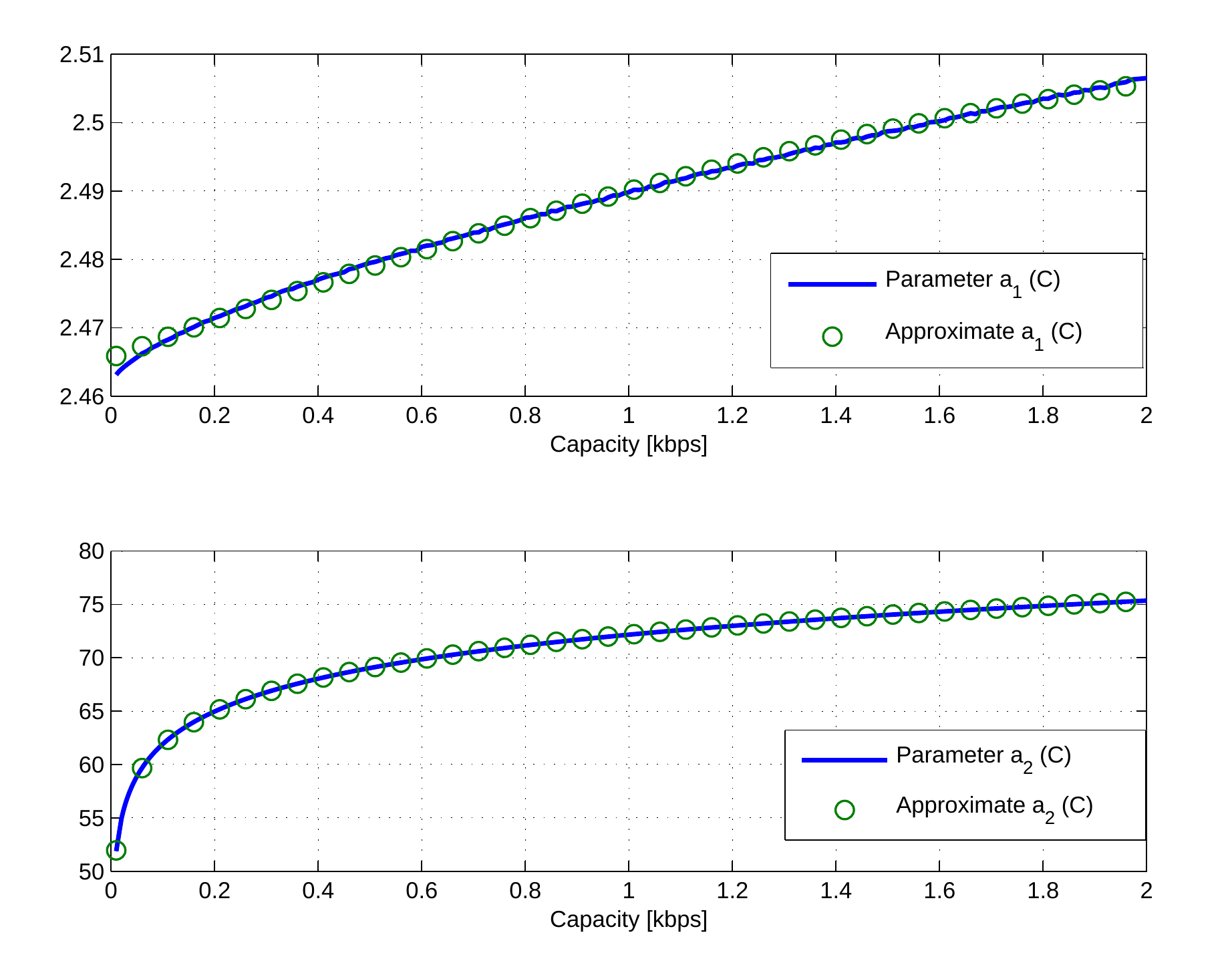}
\caption{Parameters $a_{1}$ and $a_{2}$ for $P(l,C)$ and approximate model.$l \in [0, 10km] $, $C \in [0,2kbps] $,$k=1.5$,$s = 0.5$ and $w = 0 m/s$  }
\label{figPLC_k_15_lowC.tag}
\end{figure}

\begin{figure}[t]
\centering						
\includegraphics[height=3.5in,width=3.5in]{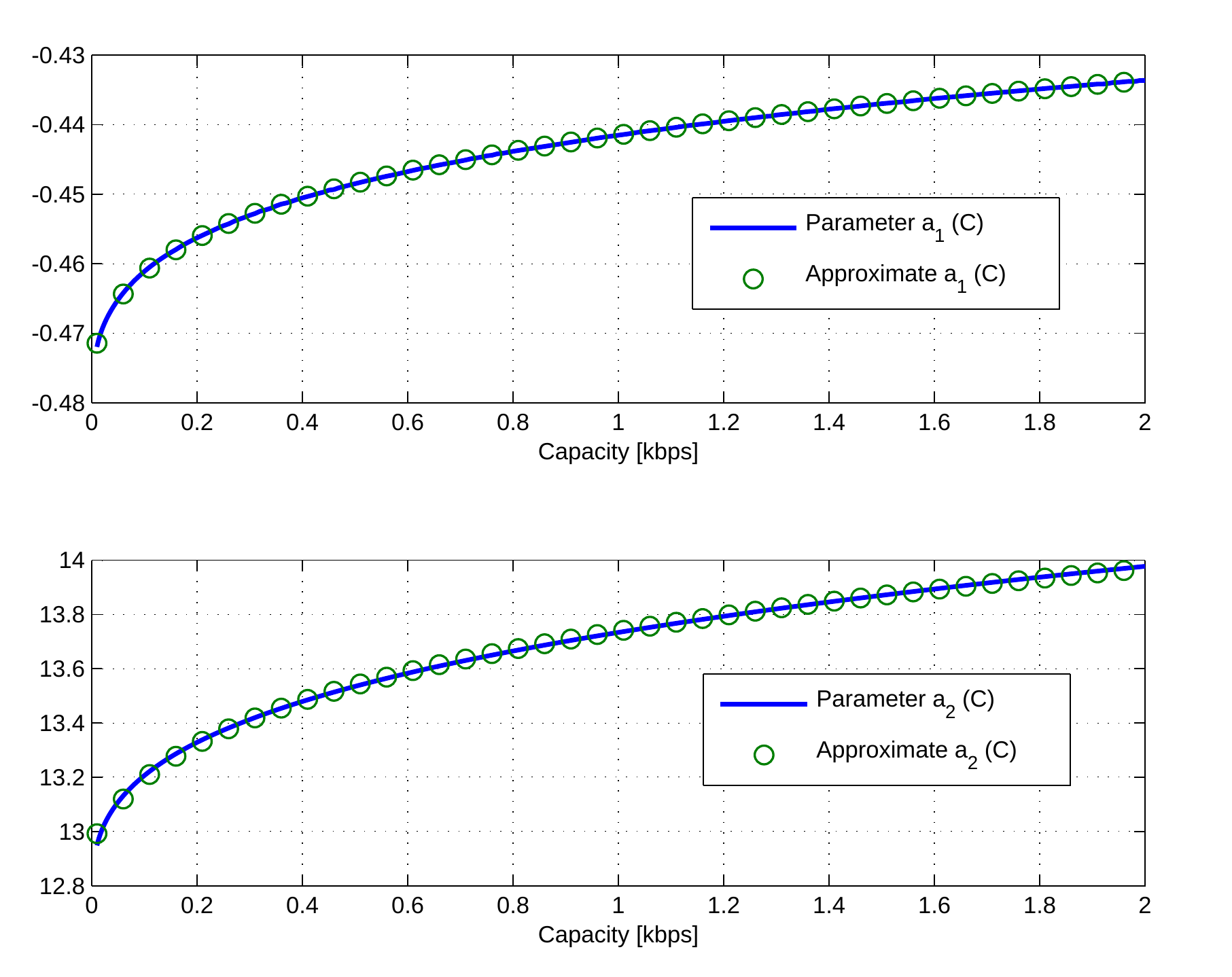}
\caption{Parameters $a_{1}$ and $a_{2}$ for $\hat{f}_{end}(l,C)$ and approximate model.$l \in [0, 10km] $, $C \in [0,2kbps]$,$k=1.5$,$s = 0.5$ and $w = 0 m/s$  }
\label{FhighLC_k_15_lowC.tag}
\end{figure}
\begin{figure}[t]
\centering	
\includegraphics[height=3.5in,width=3.5in]{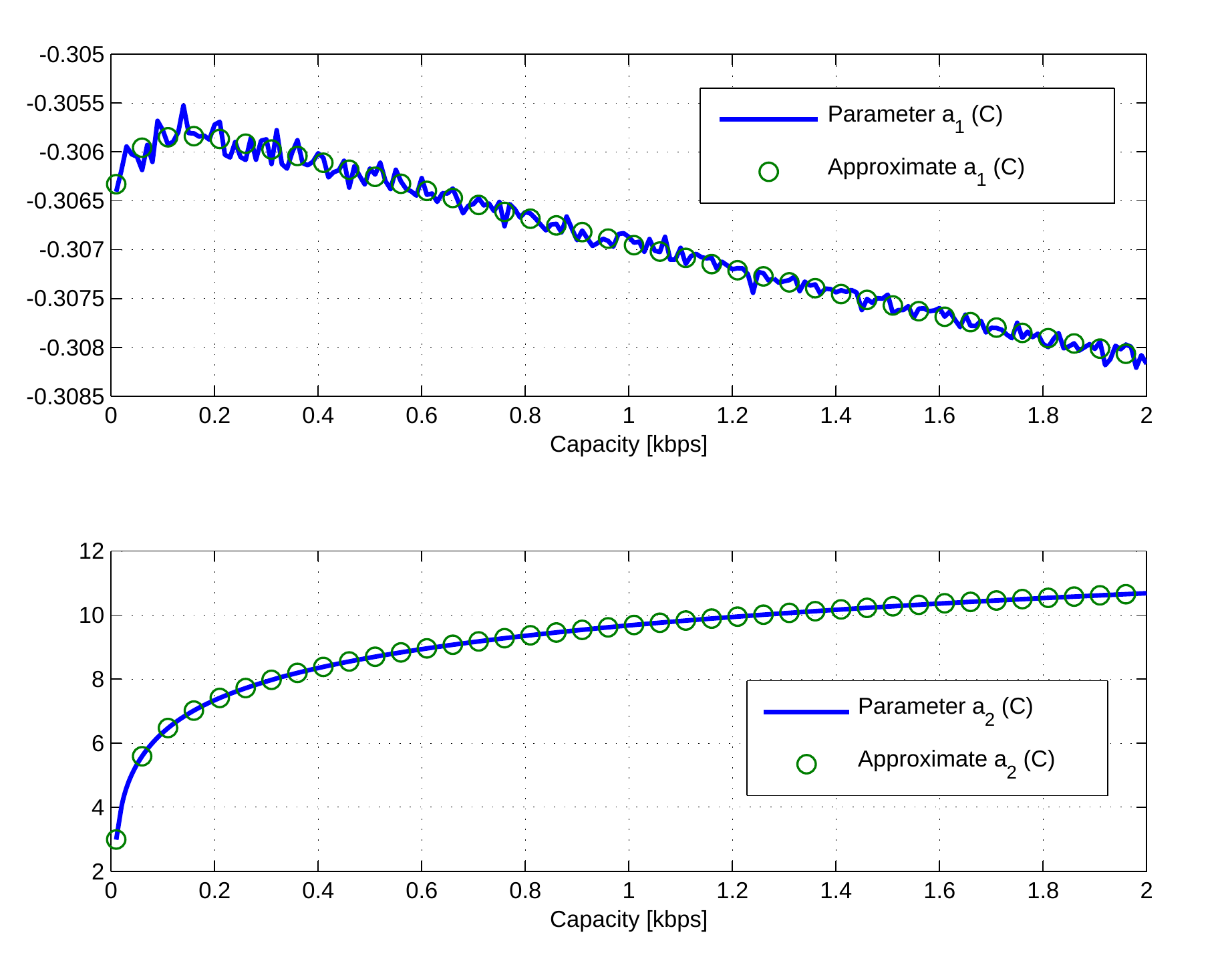}
\caption{Parameters $a_{1}$ and $a_{2}$ for $\hat{B}(l,C)$ and approximate model.$l \in [0, 10km] $, $C \in [0,2kbps]$, $k=1.5$,$s = 0.5$ and $w = 0 m/s$  }
\label{BLC_k_15_lowC.tag}
\end{figure}    	
	
\begin{figure}[t]
\centering						
\includegraphics[height=3.5in,width=3.5in]{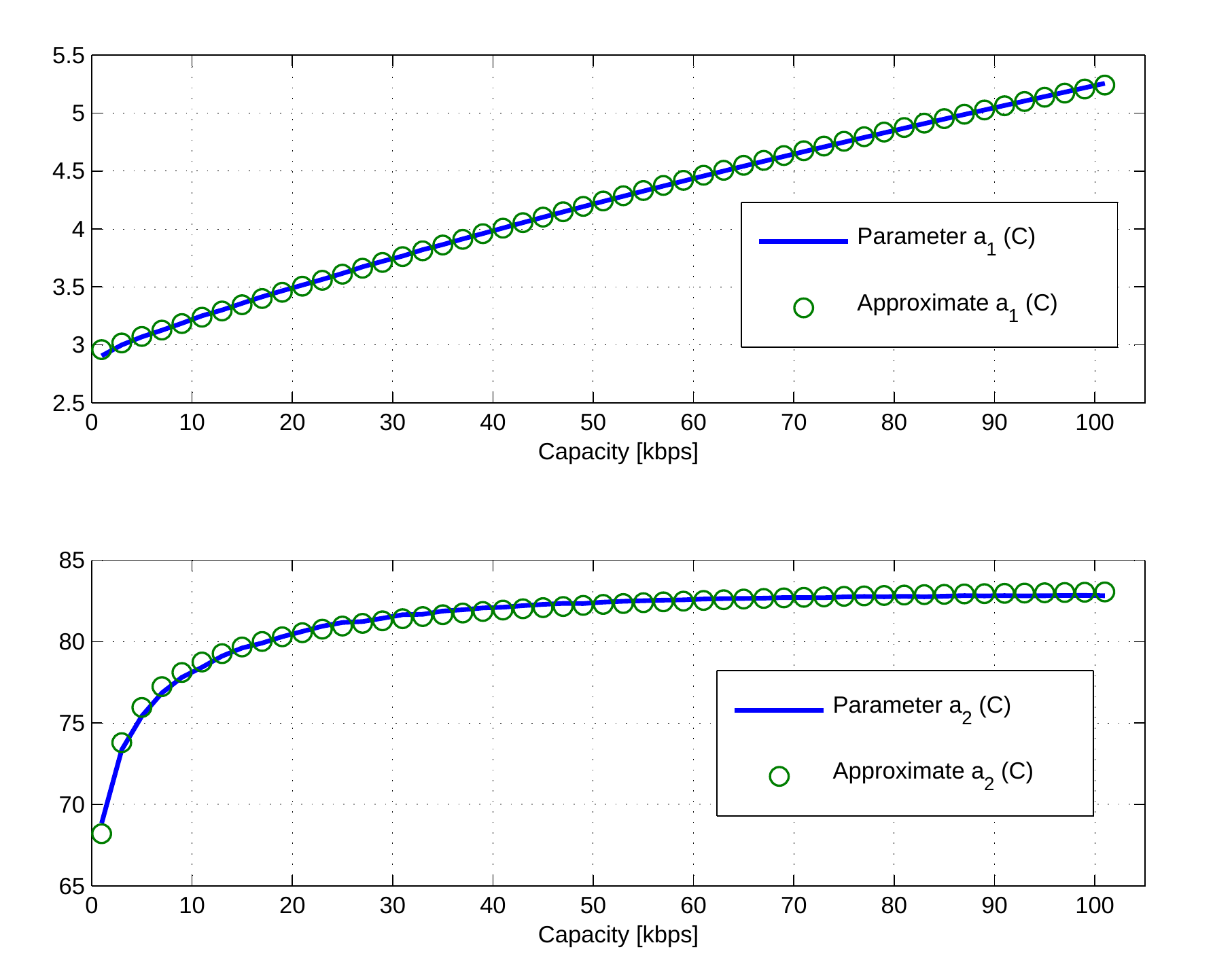}
\caption{Parameters $a_{1}$ and $a_{2}$ for $P(l,C)$ and approximate model.$l \in [0, 100km]$, $C \in [0,100kbps]$, $k=1.5$,$s = 0.5$ and $w = 0 m/s$  }
\label{figPLC_k_15_highC.tag}
\end{figure}    
	
\section{Channel Model}

An underwater acoustic channel is characterized by a path loss that depends on both distance \textit{l} and signal frequency \textit{f} as
\begin{equation}
A(l,f) = l^{k}{a(f)}^{l}
\end{equation}
where \textit{k} is the spreading factor and $a(f)$ is the absorption coefficient \cite{milica06}. The spreading factor describes the geometry of propagation, e.g. $k=2$ corresponds to spherical spreading, $k=1$ to cylindrical spreading, and $k=1.5$ to practical spreading.  The absorption coefficient can be expressed in dB/km using Thorp's empirical formula for $f$ in kHz:
\begin{align}
10\log a(f)&=0.11\frac{f^{2}}{1+f^{2}}+44\frac{f^{2}}{4100+f^{2}}\notag\\
 &+ 2.75\cdot 10^{-4}f^{2}+0.003
\end{align}
for frequencies above a few hundred Hz. For lower frequencies, the model is:
\begin{align}
10\log a(f)=0.11\frac{f^{2}}{1+f^{2}}+ 0.011f^{2}+0.002
\end{align}

The noise in an acoustic channel can be modeled through four basic sources: turbulence, shipping, waves, and thermal noise\cite{milica06}. The following formulas give the power spectral density (psd) of these noise components in dB re $\mu$ Pa per Hz as a function of frequency in kHz:
\begin{align}
&\log N_{t}(f)=1.7-3\log f\\  
&\log N_{s}(f)=4+2(s-\frac{1}{2})+2.6\log f-6\log (f+0.03)\\
&\log N_{w}(f)=5+ 0.75w^{1/2}+2\log f-4\log (f+0.4)\\
&\log N_{th}(f)=-1.5+2\log f
\end{align}
where the shipping activity $s$ ranges from 0 to 1, for low and high activity, respectively, and $w$ corresponds to the wind speed measured in m/s. The overall psd of the ambient noise is given by 
\begin{align}
N(f)&=&N_{t}(f) + N_{s}(f) + N_{w}(f) + N_{th}(f)
\end{align}

	Let us assume that this is a Gaussian channel. Then, the capacity of this channel can be obtained using the waterfilling principle \cite{milica06}. Also, assume that the power and band of operation can be adjusted to reach a certain capacity level. Thus, the capacity of a point-to-point link is

\begin{align} \label {Capacity_func}
C = \int _{B(l,C)} log_{2} \left (   \frac{K(l,C)}{A(l,f)N(f)}  \right ) df
\end{align} 

where $B(l,C)$ is the optimum band of operation. This band could be thought of as a union of non-overlapping intervals, $B(l,C) = \cup _i [f_{ini}^i(l,C),f_{end}^i(l,C)]$, where each non-overlapping band $i$ has the lower-end frequency $f_{ini}^i(l,C)$ and the higher-end frequency $f_{end}^i(l,C)$ associated to it. In its simplest form $B(l,C) = [f_{ini}(l,C),f_{end}(l,C)]$. 

The power consumption associated with a particular choice of $(l,C)$ is given by

\begin{align} \label{Power_func}
P(l,C) = \int _{B(l,C)} S(l,C,f) df
\end{align} 
	
where $ S(l,C,f) = K(l,C) - A(l,f)N(f), f \in B(l,C)$.

	Evidently, these expressions are quite complicated to be used in a computational network analysis. Also, they provide little insight into the relationship between power consumption, $\hat{f}_{ini}$ and $\hat{f}_{end}$ , in terms of the pair $(l,C)$. This motivates the need for an approximate model that will represent these relations for ranges of $C$ and $l$ that are of interest to acoustic communication systems. The model should also provide flexibility to changing other parameters, such as the spreading factor $k$, wind speed $w$ and shipping activity $s$. 

	The dependence on the spreading factor $k$ is quite simple. Let us assume that a model for $P(l,C)$ has been developed for a particular value of $k = k_{i}$, i.e. $P(l,C, k_{i})$. To determine $P(l,C, k_{j})$  for $k_{j} \neq k_{i}$, let us note that for a change in $k$, the product $A(l,f)N(f)=l^ka(f)^lN(f)$ constitutes a constant scaling factor with respect to $f$. Therefore, for a link of distance $l$ the term $B(l,C)$ will remain unchanged. Thus, if the same capacity $C$ is required for $k_{i}$ and $k_{j}$, equation \eqref{Capacity_func} shows that the only other term that can vary is $K(l,C)$, i.e. $K(l,C,k)$. Then, $K(l,C,k_{j}) = l^{k_{j} - k_{i}} K(l,C,k_{i})$. Finally, let us use the equation \eqref{Power_func} to determine the relationship between $P(l,C,k_{i})$ and $P(l,C,k_{j})$.
	The dependence on the spreading factor $k$ is quite simple. Let us assume that a model for $P(l,C)$ has been developed for a particular value of $k = k_{i}$, i.e. $P(l,C, k_{i})$. To determine $P(l,C, k_{j})$  for $k_{j} \neq k_{i}$. Note that for a change in $k$, the product $A(l,f)N(f)=l^ka(f)^lN(f)$ constitutes a constant scaling factor with respect to $f$. Therefore, for a link of distance $l$ the term $B(l,C)$ will remain unchanged. Thus, if the same capacity $C$ is required for $k_{i}$ and $k_{j}$, equation \eqref{Capacity_func}, shows that the only other term that can vary is $K(l,C)$, i.e. $K(l,C,k)$. Then, $K(l,C,k_{j}) = l^{k_{j} - k_{i}} K(l,C,k_{i})$. Finally, let us use equation \eqref{Power_func} to determine the relation between $P(l,C,k_{i})$ and $P(l,C,k_{j})$.
\begin{align}
P(l,C,k_{j}) &= \int _{B(l,C)} \left( K(l,C,k_{j}) - l^{k_{j}}a(f)^lN(f) \right ) df\\
&= l^{k_{j} - k_{i}} \int _{B(l,C)} \left(  K(l,C,k_{i}) - l^{k_{i}} a(f)^lN(f) \right ) df\\
&= l^{k_{j} - k_{i}} P(l,C,k_{i})
\end{align} 	
	
	Thus, any model for the transmission generated for some parameter $k$ has a simple extension. Also, note that the transmission bandwidth remains the same for any value of $k$.

\section{Numerical Evaluation Procedure}

\begin{figure}[t]
\centering						
\includegraphics[height=3.5in,width=3.5in]{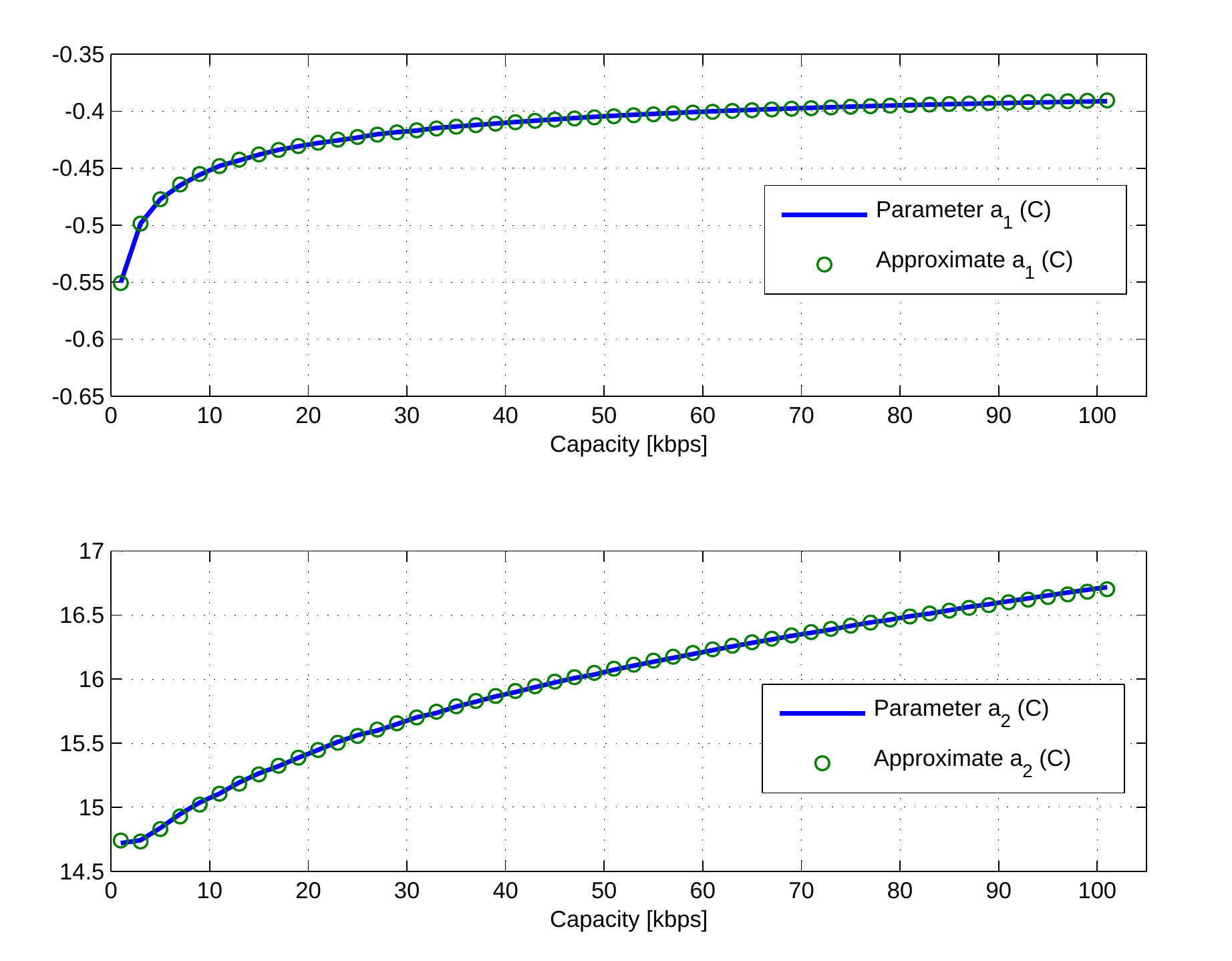}
\caption{Parameters $a_{1}$ and $a_{2}$ for $\hat{f}_{end}(l,C)$ and approximate model.$l \in [0, 100km]$, $C \in [0,100kbps] $,$k=1.5$,$s = 0.5$ and $w = 0 m/s$  }
\label{FhighLC_k_15_highC.tag}
\end{figure}
\begin{figure}[t]
\centering	
\includegraphics[height=3.5in,width=3.5in]{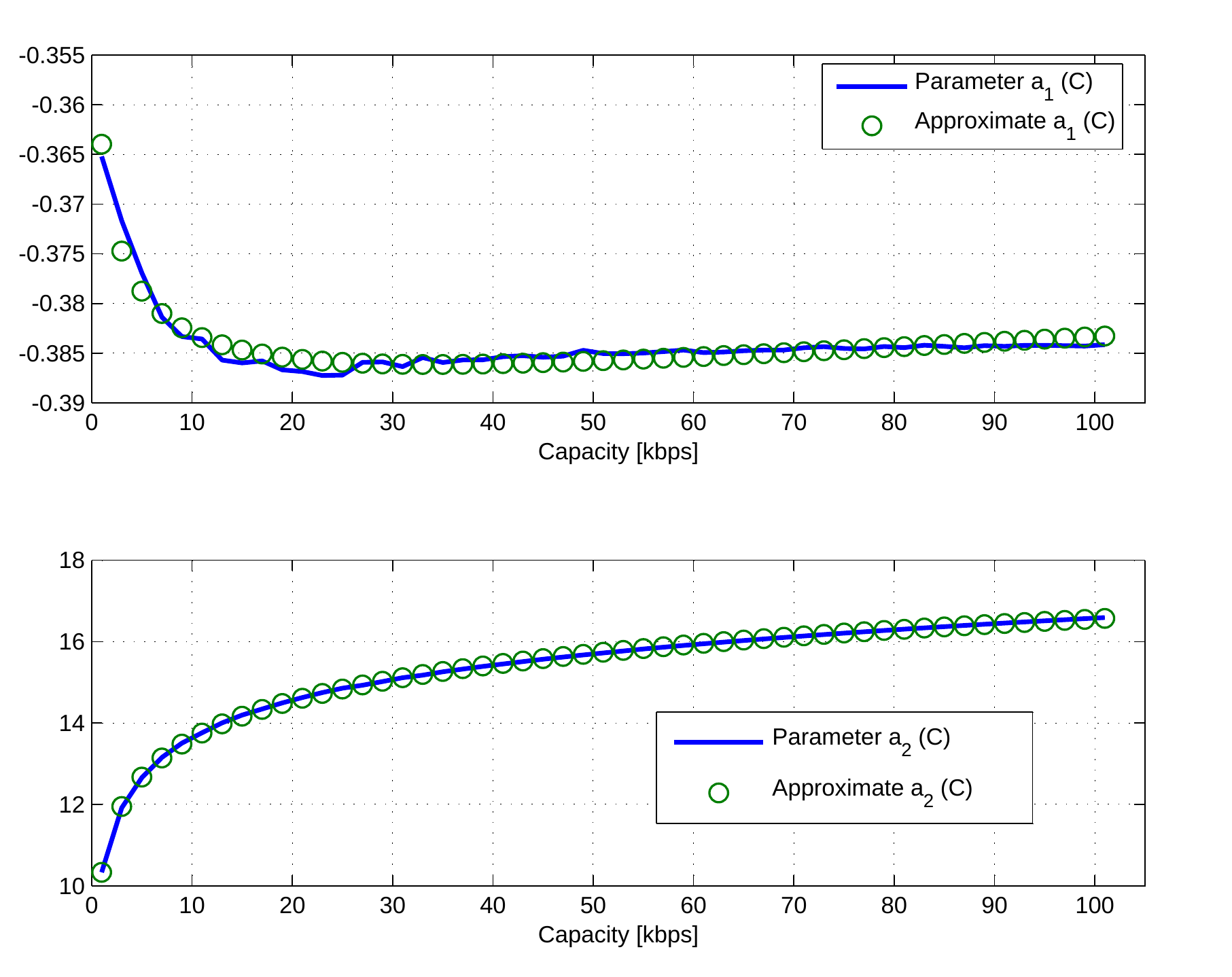}
\caption{Parameters $a_{1}$ and $a_{2}$ for $\hat{B}(l,C)$ and approximate model.$l \in [0, 100km]$, $C \in [0,100kbps]$,$k=1.5$,$s = 0.5$ and $w = 0 m/s$  }
\label{BLC_k_15_highC.tag}
\end{figure}

	A numerical evaluation procedure similar to that in \cite{milica06} is used to compute the value of $P(l,C)$, $\hat{f}_{ini}(l,C)$ and $\hat{f}_{end}(l,C)$, for a region of values of $(l,C)$. The procedure starts by fixing a target value of the capacity $C$. Then, for each distance $l$, the initial value of $K(l,C)$ is set to the minimum value of the product $A(l,f)N(f)$, i.e. $K(l,C) = \min_{f} A(l,f)N(f)$. The frequency at which this occurs, i.e. $f_{0} = \arg \min_{f} A(l,f)N(f)$, is called the optimal frequency. 

	After this, $K(l,C)$ is increased iteratively by a small amount (Figure~\ref{ANplot.tag}), until the target capacity value $C$ is met. Finally, this procedure is repeated for each value of $C$ in a range of interest.
	
	At the $n$-th step of the procedure, when $K^{(n)}(l,C)$ is increased by a small amount, the band $B^{(n)}(l,C)$ is determined for that iteration. This band is defined as the range of frequencies for which the condition $A(l,f)N(f) \leq K^{(n)}(l,C)$. Then, the capacity $C^{(n)}$ is numerically determined for the current $K^{(n)}(l,C)$ and $B^{(n)}(l,C)$, using the equation (9). If $C^{(n)} < C$, a new iteration is performed. Otherwise, the procedure stops.
 
\section{Approximate models}
	
	By applying the above procedure for varying $l$ and $C$, one arrived at the complete model for the power consumption
	
\begin{align} \label{PowerApprox}
&P(l,C) = 10^{\frac{a_{1}(C)}{10}}l^{a_{2}(C)}
\end{align} 
where

\begin{align} 
&a_{1}(C) =  \beta _3 + \beta _2 10log_{10}C  + \beta _1 (10log_{10}(C+1))^2 \label{PowerApproxA1}\\
&a_{2}(C) = \alpha _3 + \alpha _2 C  + \alpha _1 C^2 \label{PowerApproxA2}
\end{align}

Below, two ranges of operation were studied. The first one is for $l \in [0, 10] km$, $C \in [0,2] kbps$, and propagation factor of $k=1.5$, which will be called case 1 from here on. The second one is for $l \in [0, 100] Kms$, $C \in [0,100] kbps$, and propagation factor of $k=1.5$, which will be called case 2. For both regions and different ranges of $s$ and $w$, the power consumption $P(l,C)$ can be approximated by equations \eqref{PowerApprox}, \eqref{PowerApproxA1} and \eqref{PowerApproxA2}.

	Similar model are found to provide a good fit for the high/end frequency $\hat{f}_{end}(l,C)$ and for the bandwidth $\hat{B}(l,C) = \hat{f}_{end}(l,C) - \hat{f}_{ini}(l,C)$. These models are given by

\begin{align} \label{HighFreqApprox}
&\hat{f}_{end}(l,C) = 10^{\frac{a_{1}(C)}{10}}l^{a_{2}(C)}
\end{align} 
where
\begin{align} 
&a_{1}(C) =  \beta _3 + \beta _2 10log_{10}C  + \beta _1 (10log_{10}C)^2\\
&a_{2}(C) = \alpha _3 + \alpha _2 10log_{10}C  + \alpha _1 (10log_{10}C)^2
\end{align} 

\begin{align}\label{BandApprox}
&\hat{B}(l,C) = 10^{\frac{a_{1}(C)}{10}}l^{a_{2}(C)}
\end{align} 
where
\begin{align}
&a_{1}(C) =  \beta _3 + \beta _2 10log_{10}C  + \beta _1 (10log_{10}C)^2\\
&a_{2}(C) = \alpha _4 + \alpha _3 10log_{10}C + \alpha _2 (10log_{10}C)^2+\alpha _1 (10log_{10}C)^3
\end{align} 

\section{Numerical Results}
	The transmission power, highest frequency and bandwidth of transmission band were computed for a variety of values of $s$, $w$ and two ranges of interest of the pair $(l,C)$, i.e. $l \in [0, 10] Kms$, $C \in [0,2] kbps$, and $l \in [0, 100] Kms$, $C \in [0,100] kbps$. The models proposed  fitted these cases quite well. Results are presented for the case of $k = 1.5$, $w = 0$ and $s = 0.5$, for both cases. Also for case 1, it will be seen that the $\alpha$ and $\beta$ parameters show almost no dependence on the shipping activity factor $s$, especially if the wind speed is $w>0$. Thus, the approximate model for this case could be simplified to only consider $w$ as part of the model, instead of the pair $(s,w)$.

\begin{figure}[t]
\centering						
\includegraphics[height=4.2in,width=3.5in]{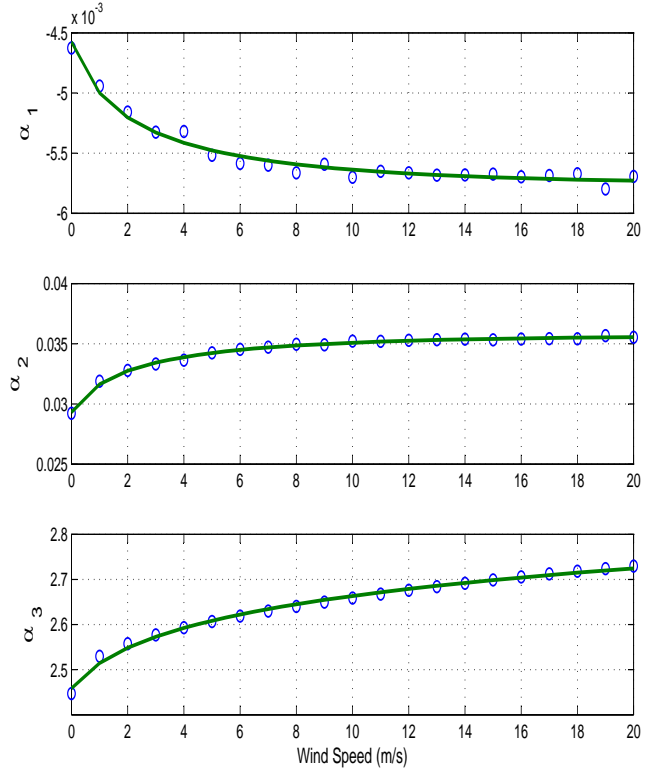}
\caption{Values of $\alpha$ as function of $w$. $l \in [0, 10km] $, $C \in [0,2kbps]$,$k=1.5$,$s = 0.5$ }
\label{PLC_k_15_lowCWindSpeed_a1.tag}
\end{figure}
\begin{figure}[t]
\centering	
\includegraphics[height=4.2in,width=3.5in]{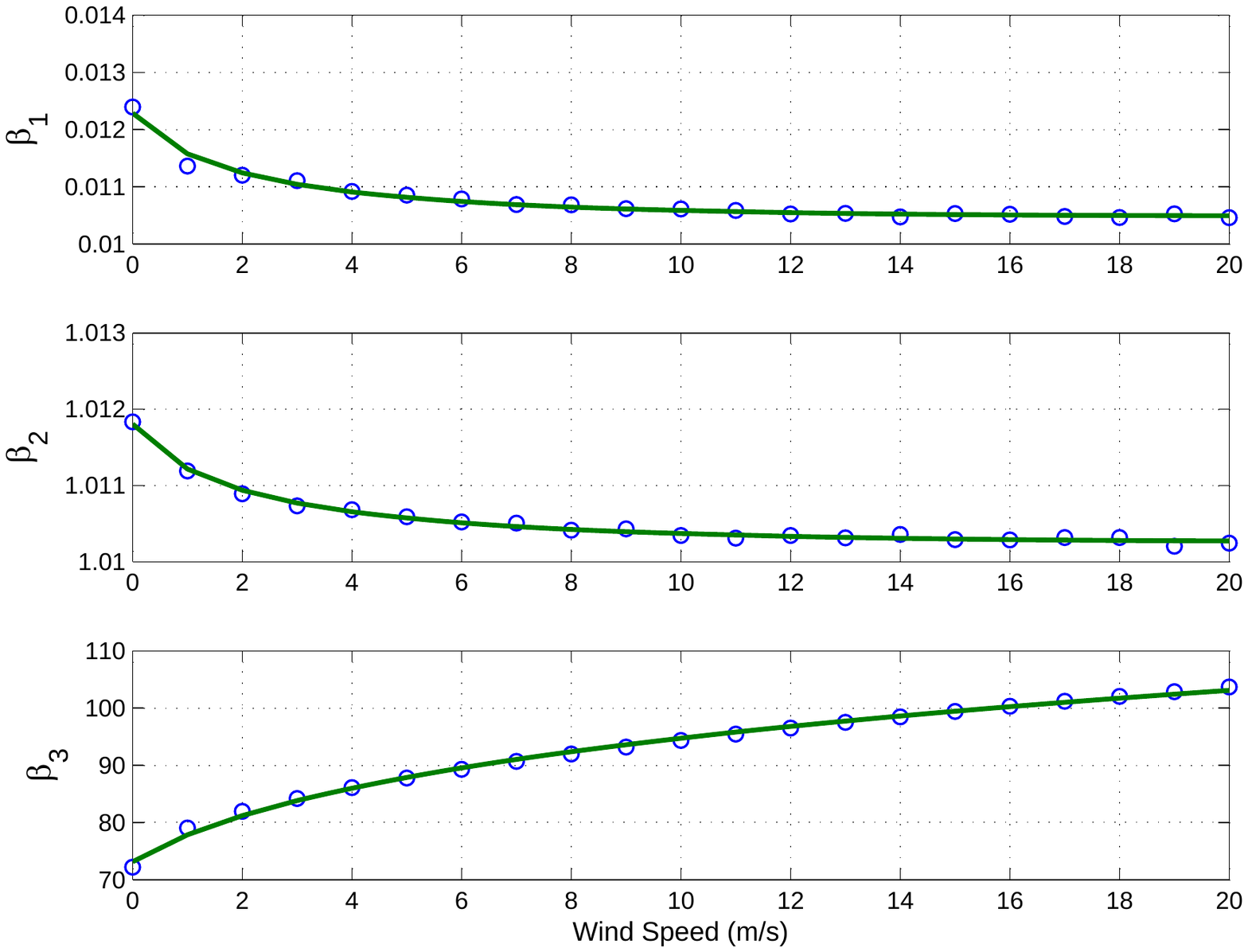}
\caption{Values of $\beta$ as function of $w$. $l \in [0, 10km] $, $C \in [0,2kbps]$,$k=1.5$,$s = 0.5$ }
\label{PLC_k_15_lowCWindSpeed_a2.tag}
\end{figure}

	Figures ~\ref{figPLC_k_15_lowC.tag}, ~\ref{FhighLC_k_15_lowC.tag}  and ~\ref{BLC_k_15_lowC.tag} show parameters $a_{1}$ and $a_{2}$ for $P(l,C)$,$\hat{f}_{end}(l,C)$,   and $\hat{B}(l,C)$, respectively. This approximation was carried out for the first case with a propagation factor of $k=1.5$, a shipping activity of $s = 0.5$ and a wind speed of $w = 0 m/s$. The values of $\alpha$'s and $\beta$'s are shown in Table ~\ref{Table_a1_lowC} and ~\ref{Table_a2_lowC}, for parameters $a_{1}$ and $a_{2}$, respectively. These tables also show the mean square error (MSE) of the approximation with respect to the actual parameters. In Figure ~\ref{BLC_k_15_lowC.tag}, there is a considerable variation in the values of parameter $a_{1}(C)$. However, note that the y-axis of the plot shows very little variation.

\begin{table}
\caption{$a_{1}$ approximation parameter values for $P(l,C)$, $\hat{f}_{end}(l,C)$ and $\hat{B}(l,C)$, with $l \in [0, 10km]$, $C \in [0,2kbps]$,$k=1.5$,$s = 0.5$ and $w = 0 m/s$  }
\centering
\label{Table_a1_lowC}
\begin{tabular}{|c||c|c|c|c|c|}
\hline
 &$\alpha _1$&  $\alpha _2$    &   $\alpha _3$  & $\alpha _4$&MSE\\
\hline
$P(l,C)$   & 0 & -0.00432 & 0.02873 & 2.46560 & 2.532e-7 \\
\hline
$\hat{f}_{end}(l,C)$   &0& 4.795e-5 &0.00246&-0.44149 & 3.930e-9 \\
\hline
$\hat{B}(l,C)$  &-5.958e-7 & -2.563e-5 & -0.000305 & -0.30694  & 6.599e-9  \\
\hline
\end{tabular}				
\end{table}
\begin{table}
\caption{$a_{2}$ approximation parameter values for $P(l,C)$, $\hat{f}_{end}(l,C)$ and $\hat{B}(l,C)$, with $l \in [0, 10km] $, $C \in [0,2kbps]$,$k=1.5$,$s = 0.5$ and $w = 0 m/s$  }
\centering
\label{Table_a2_lowC}
\begin{tabular}{|c||c|c|c|c|}
	\hline
 &  $\beta _1$    &   $\beta _2$  & $\beta _3$&MSE\\
	\hline
$P(l,C)$   & 0.01166 & 1.0117 & 72.043 & 5.8979e-5\\
\hline
$\hat{f}_{end}(l,C)$   & 0.00171 & 0.07153 & 13.738 &3.4706e-5  \\
\hline
$\hat{B}(l,C)$   & -5.163e-6 & 0.33427 & 9.6752 & 2.9233e-7 \\
\hline
\end{tabular}				
\end{table}

\begin{table}
\caption{$a_{1}$ approximation parameter values for $P(l,C)$, $\hat{f}_{end}(l,C)$ and $\hat{B}(l,C)$, with $l \in [0, 100km] $, $C \in [0,100kbps] $,$k=1.5$,$s = 0.5$ and $w = 0 m/s$  }
\centering
\label{Table_a1_highC}
\begin{tabular}{|c||c|c|c|c|c|}
\hline
 &$\alpha _1$&  $\alpha _2$    &   $\alpha _3$  & $\alpha _4$&MSE\\
\hline
$P(l,C)$   & 0 & -5.617e-5 & 0.02855 & 2.9305& 0.00011\\
\hline
$\hat{f}_{end}(l,C)$   &0& -0.00019 &0.01186 &-0.55076 & 1.32e-7\\
\hline
$\hat{B}(l,C)$  &1.696e-6 & 4.252e-5 & -0.00249 & -0.36397 &7.29e-7 \\
\hline
\end{tabular}				
\end{table}
\begin{table}
\caption{$a_{2}$ approximation parameter values for $P(l,C)$, $\hat{f}_{end}(l,C)$ and $\hat{B}(l,C)$, with $l \in [0, 100km] $, $C \in [0,100kbps]$,$k=1.5$,$s = 0.5$ and $w = 0 m/s$  }
\centering
\label{Table_a2_highC}
\begin{tabular}{|c||c|c|c|c|}
	\hline
 &  $\beta _1$    &   $\beta _2$  & $\beta _3$&MSE\\
	\hline
$P(l,C)$   & -0.032936 & 1.4104 & 67.946 & 0.04493\\
\hline
$\hat{f}_{end}(l,C)$   & 0.0065157 & -0.032693 & 14.739 &7.3024e-5 \\
\hline
$\hat{B}(l,C)$   & -0.0018252 & 0.34788 & 10.328 & 0.00019414\\
	\hline
\end{tabular}				
\end{table}

	Figures ~\ref{figPLC_k_15_highC.tag}, ~\ref{FhighLC_k_15_highC.tag}  and ~\ref{BLC_k_15_highC.tag} show parameters $a_{1}$ and $a_{2}$ for $P(l,C)$,$\hat{f}_{end}(l,C)$,   and $\hat{B}(l,C)$, respectively. This approximation was carried out for the second case with a propagation factor of $k=1.5$, a shipping activity of $s = 0.5$ and a wind speed of $w = 0$. The values of $\alpha$'s and $\beta$'s are shown in Table ~\ref{Table_a1_highC} and ~\ref{Table_a2_highC}, for parameters $a_{1}$ and $a_{2}$, respectively. These tables also show the mean square error (MSE) of the approximation with respect to the actual parameters. 
	
	For both ranges, the proposed models give a very good approximation to the actual numerical values. Also note that for the $a_2(C)$ parameter of $P(l,C)$, it is possible to use a linear approximation, instead of a quadratic model.
	
	Let us analyze the low range low rate for different values of $s$ and $w$. Table ~\ref{Table_PlCChangingSvsW} shows the values for $\alpha$ and $\beta$ parameters in the approximate $P(l,C)$ model for variations of $s$ and $w$. It is interesting that the parameters change very little with respect to the shipping activity $s$ while they show greater dependency on the wind speed factor. This is not unexpected. For low data rates and low ranges, the transmission band is at a high frequency (between 5 and 40 KHz) and from the noise equations of the model the shipping activity depends on the frequency as $O(f^{-3.4})$ while the wind speed factor has a dependency as $O(f^{-2})$. Thus, $w$ should have much more effect upon the parameters. 
	
	Therefore, a further approximation is to discard $s$ an consider parameters $\alpha$ and $\beta$ to be functions of $w$ only. Figure ~\ref{PLC_k_15_lowCWindSpeed_a2.tag} shows these relations when computed for this region a $k = 1.5$ and $s = 0.5$, which have a very simple approximation. For example, the model in equation ~\ref{EqWindSpeedModel.tag} gives a good approximation.
	
\begin{align}
\label {EqWindSpeedModel.tag}
 \psi_i(w) =  \gamma _3 + \gamma _2 10log_{10}(w+1)  + \gamma _1 (10log_{10}(w+1))^2
\end{align}

where $\psi_i(w) = \beta_i, \forall i$ and  $\psi_i(w) = \alpha_i, \forall i$. Table ~\ref{Table_windspeedrelation_lowC} shows $\gamma$ parameters for the different $alpha$'s and $\beta$'s.
	
\begin{table}
\caption{Dependency on $s$ and $w$ of $\alpha$ and $\beta$ parameters, with $l \in [0, 10km]$, $C \in [0,2kbps]$,$k=1.5$}
\centering
\label{Table_PlCChangingSvsW}
\begin{tabular}{|c|c|c|c|c|c|c|c|}
\hline
$w$& $s$ &  $\alpha _1$    &   $\alpha _2$  & $\alpha _3$ & $\beta _1$    &   $\beta _2$  & $\beta _3$ \\
\hline
  &  0   & -0.0050 & 0.0299 & 2.444  & 0.01237 & 1.0118 & 72.182\\
0 &  0.5 & -0.0046 & 0.0292 & 2.447  & 0.01239 & 1.0118 & 72.190\\
  &  1   & -0.0049 & 0.0304 & 2.469  & 0.01236 & 1.0119 & 72.271\\
\hline
  &  0   & -0.0053 & 0.0328 & 2.5572 & 0.01117 & 1.0109 & 81.960\\
2 &  0.5 & -0.0052 & 0.0327 & 2.5574 & 0.01120 & 1.0109 & 81.961\\
  &  1   & -0.0056 & 0.0336 & 2.5594 & 0.01133 & 1.0108 & 81.966\\
\hline
  &  0   & -0.0056 & 0.0342 & 2.6065 & 0.01083 & 1.0106 & 87.777 \\
5 &  0.5 & -0.0055 & 0.0342 & 2.6066 & 0.01085 & 1.0106 & 87.777\\
  &  1   & -0.0053 & 0.0339 & 2.6071 & 0.01088 & 1.0105 & 87.778\\
\hline
  &  0   & -0.0057 & 0.0351 & 2.6588 & 0.01061 & 1.0103 & 94.360\\
10&  0.5 & -0.0057 & 0.0352 & 2.6588 & 0.01061 & 1.0104 & 94.360\\
  &  1   & -0.0056 & 0.0351 & 2.6589 & 0.01059 & 1.0104 & 94.360\\
\hline
  &  0   & -0.0083 & 0.0420 & 2.7245 & 0.01079 & 1.0094 & 103.69\\
20&  0.5 & -0.0057 & 0.0355 & 2.7295 & 0.01047 & 1.0102 & 103.70\\
  &  1   & -0.0057 & 0.0355 & 2.7294 & 0.01046 & 1.0102 & 103.70\\
\hline
\end{tabular}				
\end{table}

\begin{table}
\caption{Approximation parameters of $\alpha$ and $\beta$ for $P(l,C)$, with $l \in [0, 10km] $, $C \in [0,2kbp]$, $k=1.5$,$s = 0.5$}
\centering
\label{Table_windspeedrelation_lowC}
\begin{tabular}{|c||c|c|c|}
\hline
 &  $\gamma _1$    &   $\gamma _2$  & $\gamma _3$\\
\hline
$\alpha_1$   & 5.2669e-6 & -0.000157 & -0.004575 \\
\hline
$\alpha_2$   & -2.971e-5 & 0.000865 & 0.029306 \\
\hline
$\alpha_3$   & 0.000152 & 0.01809 & 2.4586\\
\hline
$\beta_1$   & 9.924e-6 & -0.00027 & 0.012288 \\
\hline
$\beta_2$   & 7.799e-6 & -0.000219 & 1.0118 \\
\hline
$\beta_3$   & 0.068091 & 1.3659 & 73.144\\
\hline
\end{tabular}				
\end{table}


%

%

\section{Conclusion}

	This paper offers an insight into the dependence of the transmission power, bandwidth, and the band-edge frequency of an underwater acoustic link on the capacity and distance. It provides closed-form approximate models for the time-invariant acoustic channel, taking into account a physical model of acoustic path loss and the ambient noise, assuming that the channel is Gaussian. These approximate models where shown to provide a good fit to the actual empirical values by numerical evaluation for different ranges of distance $l$ and capacity $C$, as well as noise profiles corresponding to different shipping activity factor and wind speed. 

	The band-edge frequency $\hat{f}_{end}(l,C)$ and the bandwidth $\hat{B}(l,C)$ were also shown to be invariant to the spreading factor $k$, while the power scales as $P(l,C,k') = l^{k' - k} P(l,C,k)$. 

	For a certain range of values (l,C), the approximate model of $P(l,C)$ was shown to be almost independent of the shipping activity factor $s$ while having a marked dependency on the wind speed $w$. This dependence, however, is quite smooth and could be approximated by a simple model, thus resulting in a complete model for the $P(l,C)$ for a range of values (l,C) that is of interest to a typical underwater communciation system. Hence, these models can be used in network optimization problems to determine the optimal power consumption for some required data rate.

	Future work will focus on studying convexity properties of the $P(l,C)$ model and using it in network optimization problems.


\section*{Acknowledgment}
This work was supported in part by the NSF grants \# 0520075 and ONR MURI Grant \# N00014-07-1-0738, and DARPA BAE Systems National Security Solution, Inc. subcontract \# 060786.

\end{document}